\documentclass[fleqn,usenatbib]{mnras}

\usepackage{newtxmath}
\usepackage[T1]{fontenc}
\usepackage{ae,aecompl}

\usepackage{graphicx}	% Including figure files
\usepackage{amsmath}	% Advanced maths commands
\usepackage{amssymb}	% Extra maths symbols

\newcommand{\pc}{\,\mathrm{pc}}

\title[The Force of a Dwarf]{Kinematics with \textit{Gaia} DR2: The Force of a Dwarf}
\author[I. Carrillo et al.]{
I. Carrillo,$^{1}$\thanks{E-mail: icarrillo@aip.de}
I. Minchev,$^{1}$
M. Steinmetz,$^{1}$
G. Monari,$^{1}$
C. F. P. Laporte,$^{2}$\thanks{CITA National Fellow}
\newauthor
F. Anders,$^{1}$
A. B. A. Queiroz,$^{1}$
C. Chiappini,$^{1}$
A. Khalatyan,$^{1}$
M. Martig, $^{3}$
\newauthor
P. J. McMillan,$^{4}$
B. X. Santiago,$^{5,6}$
K. Youakim$^{1}$
\\
$^{1}$Leibniz Institut f\"{u}r Astrophysik Potsdam (AIP), An der Sterwarte 16, D-14482 Potsdam, Germany\\
$^{2}$Department of Physics $\&$ Astronomy, University of Victoria, 3800 Finnerty Road, Victoria BC, Canada V8P 5C2\\
$^{3}$Astrophysics Research Institute, Liverpool John Moores University, 146 Brownlow Hill, Liverpool L3 5RF, UK\\
$^{4}$Lund Observatory, Lund University, Department of Astronomy and Theoretical Physics, Box 43, SE-22100, Lund, Sweden\\
$^{5}$Instituto de F\'{i}sica, Universidade Federal do Rio Grande do Sul, Caixa Postal 15051,Porto Alegre, RS - 91501-970, Brazil\\
$^{6}$Laborat\'{o}rio Interinstitucional de e-Astronomia, - LIneA, Rua Gal. Jos\'{e} Cristino 77, Rio de Janeiro, RJ - 20921-400, Brazil}
\date{Accepted XXX. Received YYY; in original form ZZZ}

\pubyear{2018}
\begin{document}
\label{firstpage}
\pagerange{\pageref{firstpage}--\pageref{lastpage}}
\maketitle

% Abstract of the paper
\begin{abstract}
We use {\it Gaia} DR2 astrometric and line-of-sight velocity information combined with two sets of distances obtained with a Bayesian inference method to study the 3D velocity distribution in the Milky Way disc. We search for variations in all Galactocentric cylindrical velocity components ($V_{\phi}$, $V_R$ and $V_z$) with Galactic radius, azimuth, and distance from the disc mid-plane. We confirm recent work showing that bulk vertical motions in the $R\text{-}z$ plane are consistent with a combination of breathing and bending modes. In the $x\text{-}y$ plane, we show that, although the amplitudes change, the structure produced by these modes is mostly invariant as a function of distance from the plane. 
Comparing to two different Galactic disc models, we demonstrate that the observed patterns can drastically change in short time intervals, showing the complexity of understanding the origin of vertical perturbations.
A strong radial $V_R$ gradient was identified in the inner disc, transitioning smoothly from $16$\,km\,s$^{-1}$\,kpc$^{-1}$ at an azimuth of $30^\circ<\phi<45^\circ$ ahead of the Sun-Galactic centre line, to $-16$\,km\,s$^{-1}$\,kpc$^{-1}$ at an azimuth of $-45^\circ<\phi<-30^\circ$ lagging the solar azimuth. We use a simulation with no significant recent mergers to show that exactly the opposite trend is expected from a barred potential, but overestimated distances can flip this trend to match the data. Alternatively, using an $N$-body simulation of the Sagittarius dwarf-Milky Way interaction, we demonstrate that a major recent perturbation is necessary to reproduce the observations. Such an impact may have strongly perturbed the existing bar or even triggered its formation in the last $1\text{-}2$\,Gyr.
\end{abstract}
% Select between one and six entries from the list of approved keywords.
% Don't make up new ones.

%%%%%%%%%%%%%%%%%%%%%%%%%%%%%%%%%%%%%%%%%%%%%%%%%%

%%%%%%%%%%%%%%%%% BODY OF PAPER %%%%%%%%%%%%%%%%%%
\defcitealias{Carrillo2018MNRAS}{C18}

\section{Introduction}
It is now well established that the Milky Way disc contains non-axisymmetric structures, such as the Galactic bar and spiral arms. The effect of these asymmetries has been linked to substructure in the local stellar phase-space, first seen clearly as clumps in the velocity distribution \citep{1998Dehnen} using data from the Hipparcos astrometric satellite \citep{hipparcos}. Using only this small disc patch around the Sun ($d< 200$~pc), it was already possible to establish the need for spiral (\citealp{Quillenspiral}; \citealp{Pompeia2011}) and bar structure (\citealp{2000Dehnen}; \citealp{Minchev2007}; \citealp{2009Antoja}). 

Pre-{\it Gaia}, non-zero mean motions were found in the radial direction (e.g. \citealp{siebertgradient}; \citealp{wobbly}) with origins usually attributed to internal perturbations due to the bar and spiral arms (e.g. \citealp{Siebert2012}; \citealp{Monari2014}). Non-axisymmetries were also found in the direction perpendicular to the Galactic disc (\citealp{Widrow2012}; \citealp{wobbly}; \citealp{Carlin2013}) whose origins are still under debate and may result from both internal (e.g. \citealp{2014Faure}; \citealp{Monari2015}; \citealp{2016Monari}) and external perturbations, such as the passage of a satellite galaxy or a dark matter subhalo (\citealp{Gomez2013}; \citealp{Widrow2014}; \citealp{Elena2016}; \citealp{Laporte2017}).

The emerging {\it Gaia} era, where distances and proper motions are being obtained with unprecedented number, accuracy and precision, is changing our understanding of the velocity distribution in the Milky Way. Thanks to these data, we are now in a much better position to trace the origins of disc asymmetries. These features could be used as a dynamical diagnostic to model the history of interactions between the Milky Way and its satellite galaxies, as well as understand the importance of internal mechanisms.

Already with just early mission data from the first data release {\it Gaia} DR1 \citep{GaiaDR1first} combined with data from the Radial Velocity Experiment (RAVE; \citealp{DR1}) fifth data release (DR5; \citealp{RAVEDR5}) and inferred distances by \citet{Astraatmadja} and \citet{McMillan2017}, \citet[][hereafter C18]{Carrillo2018MNRAS} studied the 3D velocity distribution in the extended solar neighbourhood, focusing on north-south differences and the observed vertical velocity asymmetries. In contrast to previous work showing a rarefaction-compression behaviour, in which the vertical velocity distribution has odd parity with respect to the Galactic plane and even parity in the density distribution, known as a breathing mode, \citetalias{Carrillo2018MNRAS} showed a vertical pattern with a combination of a breathing mode inside the solar radius ($R_0$) and a bending mode (even parity in $V_z$ with odd parity in the density distribution) outside $R_0$. This mode combination is physically intuitive as inwards of the solar radius the Galactic bar and spiral arms naturally induce breathing modes (\citealp{2014Faure}; \citealp{Monari2015}; \citealp{2016Monari}), while outside $R_0$ bending modes are consistent with external perturbations such as a passing satellite galaxy or, on a different scale, dark matter subhaloes. \citetalias{Carrillo2018MNRAS} also showed that $V_z$ depends strongly on the adopted proper motions and that distance uncertainties can create artificial wave-like patterns.

The accuracy and precision achieved in the second {\it Gaia} data release ({\it Gaia} DR2) has brought a significant increase of scientific discoveries. A considerable effort mapping the kinematics of {\it Gaia} DR2 was made by \citet{Katz2018} finding a variety of velocity substructures and seeing previously known ones with much more clarity. In comparison to the combination of breathing and bending mode observed by \citetalias{Carrillo2018MNRAS} with {\it Gaia} DR1, \citet{Katz2018} showed that as the volume around the Sun extended to larger distances the bending mode outside $R_0$ reversed direction (as previously seen in the outer disc using LAMOST \citep{Wang2018}), while a breathing mode inside $R_0$ was confirmed.

\citet{Katz2018} and \citet{Ramos2018} showed that the solar neighborhood in-plane velocity distribution (known as the $U-V$ plane, with $U$ and $V$ the Galactocentric radial and tangential stellar velocity components) exhibits long arches separated by $\sim20$\,km\,s$^{-1}$ in $V$, in addition to the well known low-velocity moving groups (such as Hercules, Hyades, etc.). This confirmed predictions by \cite{Minchev2009}, who used a semi-analytical model and test-particle simulations to show that phase wrapping following a perturbation from a recent merger could create such arches, linking the separation between them to an impact $\sim2$~Gyr ago. Some of the arches found in {\it Gaia} DR2 $U-V$ plane could also be interpreted as the effect of spiral arm crossings \citep{Quillen2018}, transient spiral arms \citep{Hunt2018}, and the effect of a long bar \citep{Monari18}.

Another relevant discovery using {\it Gaia} DR2 data was the phase-space ($z, V_z$) spiral found in the distribution of solar neighbourhood radial and azimuthal velocities by \citet{Antoja2018}, interpreted as further evidence for ongoing phase mixing in the Milky Way disc, due to the Sagittarius dwarf galaxy \citep[see e.g.][]{Johnston2005, Law2010, Laporte2017}. This result reinforced earlier interpretations of observational signs in the Milky Way taken as evidence that the Galaxy was perturbed by external agents \citep[see e.g.][]{Minchev2009, Kazantzidis2009, Gomez2012b,Elena2016} and in particular Sagittarius \citep[][]{Quillen2009, Purcell2011, Gomez2013, delaVega2015}.

Although phase-mixing neglects self-gravity and structure dissipates with time, the patterns reproduced by \citet{Laporte2018} obtained pre-{\it Gaia} DR2 using a full $N$-body simulation of the impact of Sagittarius with the Milky Way are consistent with the timescale and structure seen by \citet{Antoja2018}. \citet{Binney2018} explained the phase-space spiral using toy-models of tracers defined by distribution functions reacting to a point mass perturber. \citet{Chequers2018} used $N$-body simulations of an isolated disc-bulge-halo system to investigate the continual generation of bending waves by a system of satellites or dark matter subhaloes. Their results suggest that the phase-space spiral may be due to long-lived waves of a continually perturbed disc rather than, or in addition to, single/recent satellite interactions. Similarly, \citet{Darling2018} used test-particles simulations and $N$-body models to show that disc bending perturbations can naturally create phase-space spirals. \citet{BlandHawthorn2019} used {\it Gaia} DR2 data complemented with spectroscopy from the GALAH survey (\citealp{DeSilva2015}; \citealp{Martell2017}) to study the phase-space spiral in abundance and action spaces. On the other hand, \citet{Khoperskov2018} used a high-resolution $N$-body simulation to demonstrate that not only external perturbations can create structures qualitatively similar to those observed by \citet{Antoja2018}. They showed that vertical oscillations driven by the buckling instability of the bar could also naturally create phase-space spirals. The different approaches used to explain the origins of the phase-space spiral suggests that this structure might result from a superposition of waves due to internal and external mechanisms and disc phase-wrapping.

{\it Gaia} DR2 has also made it possible to identify a number of ridges of negative slopes in the $V_\phi-R$ stellar distribution \citep{ Antoja2018, Kawata2018, Laporte2018, Fragkoudi2019, Khanna2019} separated roughly by $20-30$\,km\,s$^{-1}$, similarly to the arches found in the $U-V$ plane. This structure has been interpreted as phase-wrapping due to a perturbation with Sagittarius \citep{Antoja2018}, internal spiral arms \citep{Kawata2018, Hunt2018}, or the Galactic bar \citep{Fragkoudi2019}. Note that a ridge interpreted as the bar's Outer Lindblad resonance in the $V_\phi-R$ plane was first identified in {\it Gaia} DR1 data by \cite{Monari2017}.

In this work we follow up on the analysis of \citetalias{Carrillo2018MNRAS} using data from the second {\it Gaia} data release ({\it Gaia} DR2; \citealp{Gaiadr2}) and two sets of distances obtained via Bayesian inference. The used priors consider the parallax and optical photometry from {\it Gaia} as well as multiband photometric information. These inferred distances, together with the outstanding improvement compared to the proper motions of {\it Gaia} DR1, allows for a deeper study of the radial and vertical motions by expanding the volume observed by \citet{Katz2018}. This could help us to further constrain the origins of such perturbations.

This paper is structured as follows: In Section~\ref{sec:data_coord}, we introduce the data, our sample selection and present a comparison of our inferred distances and the direct use of the inverse parallax. A brief description of the Galactic disc models used in this work is given in Section~\ref{sec:Models}. In Section~\ref{sec:3dVel}, we present our kinematic analysis of the radial, azimuthal and vertical velocity distribution as a function of radius ($R$), azimuth ($\phi$), and height ($z$). Here we also briefly show the effects of systematic errors in the data and we further study the radial and vertical velocity distributions compared to simulations. Finally, Section~\ref{sec:Concl} contains a summary of our results.
\section{Data}
\label{sec:data_coord}

\subsection{Coordinate system} 
\label{sec:cylindrical} 

We used the sky positions, line-of-sight velocities, and proper motions from {\it Gaia} DR2, as well as two sets of distance estimates presented in Section~\ref{sec:Distances}, to compute the Galactocentric positions and velocities for stars in our sample. The method used to derive the velocities is described in detail by \citet{1987}. We computed the Galactocentric cylindrical velocity components ($V_R$, $V_{\phi}$ and $V_z$), following the coordinate transformation given in Appendix A of \citet{wobbly}.

We use the estimate of the peculiar velocity of the Sun obtained by \citet{binney}:
\begin{equation}
\label{eq:Sunpeculiar}
(U,V,W)_{\odot}=(11.1,\,12.24,\,7.25)\,\text{km s}^{-1}
\end{equation}
and adopt $R_0=8.34$\,kpc \citep{Reid2014} for the solar Galactocentric distance. 
With these values and the proper motion of Sagittarius\,A*, $\mu_{l_\mathrm{
Sgr\,A^*}}\,=6.379$ mas\, yr$^{-1}$ \citep{Reid2004}, we obtain:
\begin{equation}
V_{\odot}+V_\mathrm{LSR} = 4.74\,R_0\,\mu_{l_\mathrm{ Sgr\,A^*}},
\end{equation} 
which yields a value of $V_\mathrm{LSR}\sim240$\,km\,s$^{-1}$ for the
circular velocity of the local standard of rest (LSR). The selection of these parameters will affect the $V_R$ and $V_{\phi}$ values, but since $V_z$ is independent of $V_\mathrm{LSR}$ it will suffer no change.

\subsection{\textit{Gaia} DR2 sample selection}
\label{sec:selection}

ESA's mission {\it Gaia} \citep{Gaia_mission} is acquiring highly accurate
parallaxes, proper motions, radial velocities, and astrophysical parameters for over a billion sources that will allows us to study the dynamics, the structure and the origins of our Galaxy. The current second data release {\it Gaia} DR2, contains median radial velocities for more than 7.2 million stars with a mean G magnitude between $\sim 4$ and 13 and an effective temperature in the range of $\sim 3550$ to 6900 K. The overall precision of the radial velocities (on the order of $\sim200-300$\,m\,s$^{-1}$ and $\sim1.2$\,km\,s$^{-1}$, at the bright and faint ends, respectively) represents a considerable improvement to previous data. However, the biggest improvement comes from the parallaxes and proper motions with typical uncertainties of 0.04\,mas and 0.06\,mas\,yr$^{-1}$, respectively. This will improve the velocity uncertainties up to a factor of $6$ compared to the results presented in \citetalias{Carrillo2018MNRAS}.

In this work we select our {\it Gaia} DR2 sample based mainly on the recommended astrometric quality indicators such as the unit weight error (UWE) and the renormalised unit weight error (RUWE). The RUWE is computed following the recipe of \citet{Lindegren2018}:
\begin{equation}
\mathrm{\mathrm{RUWE=UWE}}/u_{0}(G,C)
\end{equation}
where ,
\begin{equation}
\mathrm{UWE}=\sqrt{\chi^{2}/(N-5)}
\end{equation}
\begin{equation*}
\chi^{2}=\texttt{astrometric\_chi2\_al}
\end{equation*}
\begin{equation*}
N=\texttt{astrometric\_n\_good\_obs\_al},
\end{equation*}
\begin{equation*}
G=\texttt{phot\_g\_mean\_mag},
\end{equation*}
\begin{equation*}
C= \texttt{phot\_bp\_mean\_mag\,-\,phot\_rp\_mean\_mag}.
\end{equation*}

$u_{0}(G,C)$ is a normalisation factor obtained from interpolating in $G$ and $C$ the provided table on the ESA {\it Gaia} DR2 \href{https://www.cosmos.esa.int/web/gaia/dr2-known-issues}{known issues page}.

We select stars with $\mathrm{RUWE}\leq1.4$, which ensures astrometrically well-behaved sources. Further cuts include to select stars with \texttt{VISIBILITY\_PERIODS\_USED}\,$>8$, these cuts select stars with a better parallax determination and remove stars with parallaxes more vulnerable to errors. Finally, we select stars with an inferred distance uncertainty of $\sigma_{d}/d<0.2$.  
\subsection{Distance estimates}\label{sec:distance}
\label{sec:Distances}

\begin{figure*}
	\includegraphics[width=\textwidth]{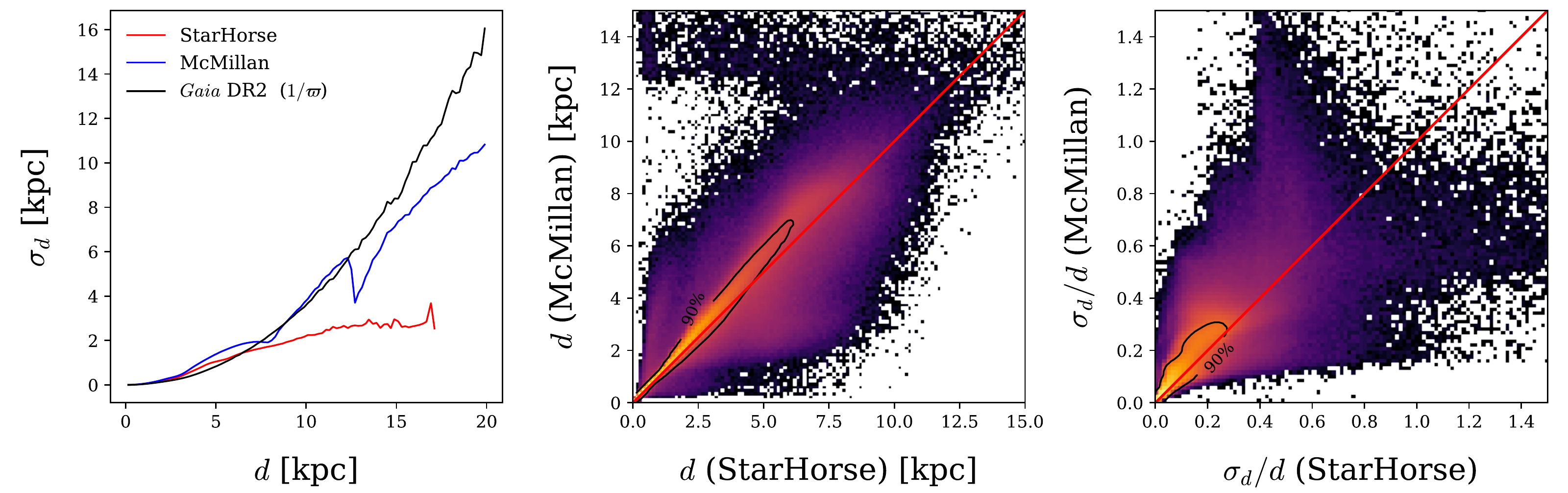}
	\caption{Comparison of inferred distances, as well as the direct use of the inverse parallax of {\it Gaia} DR2. The left panel shows the median uncertainty as a function of distance in bins of 0.2\,kpc. Here the {\tt StarHorse} estimate remains relatively constant also at large distances, while the McMillan estimate and the inverse parallax follow a similar profile. The agreement between both our distances is displayed in the middle panel, where the red line indicates a perfect match between values. The shift at distances $\gtrsim 3$\,kpc is mainly due to dust extinction being taken into account by {\tt StarHorse}. The right panel displays the relative distance uncertainties $\sigma_d/d$, showing that all distances are more precise when using the {\tt StarHorse} estimates.}
	\label{fig:distcomp}
\end{figure*}

Over the last few years, it has become clear that using the inverse parallax as a distance estimator is not trivial (e.g. \citealp{Brown1997}; \citealp{ArenouLuri1999}; \citealp{CBJ1}; \citealp{Astraatmadja}; \citealp{Luri2018}). Parallaxes containing an error larger than $20\%$ result in biased distance estimates but it is also not recommended to select only stars with lower parallax uncertainties (although widely used), as this also creates biases. Bayesian inference therefore, has become a common technique to infer parallaxes from the observed data. In this Section, we briefly describe the methods used to estimate both sets of inferred distances and present a comparison of both samples with the inverse parallax of {\it Gaia} DR2.

\citet{Paul2018} estimated stellar distances, $s$, using the observed {\it Gaia} parallax, $\varpi$, and the radial velocity spectrometer magnitude, $G_{RVS}$, (calculated using the approximation given by \citealp{Gaiadr2}). The Bayesian estimate is then given by: 
\begin{equation*}
P(s|\, \varpi,G_{RVS}) \propto \; P(\varpi | s,\sigma_\varpi) \times s^2 P(\mathbf{r}) P(M_{G_{RVS}}) ,
\end{equation*} 
where $M_{G_{RVS}}$ is the absolute magnitude in the $G_{RVS}$ band. The prior $P(M_{G_{RVS}})$, which is an approximation to the distribution of $M_{G_{RVS}}$, was modelled with PARSEC isochrones \citep{Marigo2017}.
The density model used to give $P(\mathbf{r})$ is the same as that used by \citet{McMillan2017}, i.e., a three component model (thin disc, thick disc and halo, labelled a, b and c respectively), 
\[
P(\mathbf{r}) \propto \mathrm{N_{\mathrm{a}}}\,\exp\left(-\frac{R}{R_d^{\mathrm{a}}} - \frac{|z|}{z_d^{\mathrm{a}}} \right) + \mathrm{N_{\mathrm{b}}}\,\exp\left(-\frac{R}{R_d^{\mathrm{b}}} - \frac{|z|}{z_d^{\mathrm{b}}} \right) + \mathrm{N_{c}}\, r^{-3.39}
\]
with $R_d^{\rm a}=2\,600\pc$, $z_d^{\rm a}=300\pc$, $R_d^{\rm b}=3\,600\pc$, $z_d^{\rm b}=900\pc$ and normalisations $\mathrm{N_{i}}$ (see \citealp{McMillan2017} and \citealp{Paul2018} for further details).

While no colour information nor dust extinction was considered, the obtained distance nonetheless represents an important improvement over the na\"{i}ve inverse parallax estimate.

\citet{Anders2019} derived distances using the {\tt StarHorse} code (\citealp{Santiago2016}; \citealp{Queiroz2018}), which finds the posterior probability over a grid of stellar models, distances and extinctions, given a set of astrometric, spectroscopic, and photometric observations and a number of Galactic priors. For this, they combined parallaxes, adopting a fixed parallax zero-point shift of 0.05\,mas \citep{Zinn2018} and {\it Gaia} DR2 optical photometry ($G, G_{\rm BP}, G_{\rm RP}$) with the photometric catalogues of 2MASS $JHK_s$, PANSTARRS and AllWISE $W1,W2$. Other priors used include an initial mass function as well as stellar density, metallicity, and age distributions for the main Galactic stellar components (see \citealp{Queiroz2018} for further details). Due to the multi-band photometric data, the realistic Galactic density priors used and the treatment of dust extinction, an improvement in the distance precision is expected.

The left panel of Fig.\,\ref{fig:distcomp} shows the distance uncertainty as a function of our two sets of inferred distances, as well as the distance obtained directly by computing the inverse parallax from {\it Gaia} DR2. Here it is clear, that {\tt StarHorse} distances have smaller uncertainties at large ranges. The middle panel displays a comparison between our distance estimates. Both {\tt StarHorse} and McMillan distances agree quite well up to $\approx 3$\,kpc, beyond which we see a shift towards larger distances by McMillan compared to {\tt StarHorse}. The shift may be due to dust extinction, which decreases apparent stellar brightness and if unaccounted for, leads to overestimate distances. The right panel of Fig.\,\ref{fig:distcomp} displays a comparison of the relative distance uncertainties. Here, as previously expected, we see that the distance estimates from {\tt StarHorse} have smaller uncertainties than the ones obtained by \citet{Paul2018}. 

As a result of the more precise estimates and the extinction treatment, we use the {\tt StarHorse} distances derived by \citet{Anders2019} as the main distance set in this work. The data selection described in Section~\ref{sec:selection} yields a sample of 5,167,034 stars using the inferred distances of \citet{Paul2018} and 5,420,754 stars using {\tt StarHorse}, which include additionally the cuts described in \citet{Anders2019}: $\texttt{SH\_GAIAFLAG}="000"$ and $\texttt{SH\_OUTFLAG}="00000"$.

Another approach to obtain distance estimates was recently used by \citet{Corredoira2018}, who used Lucy's inversion method of the Fredholm integral equations of the first kind to estimate distances up to 20\,kpc. This method, however, makes use of no Galaxy priors and does not consider extinction, which as discussed, overestimate distances. 

\begin{figure*}
	\includegraphics[width=\textwidth]{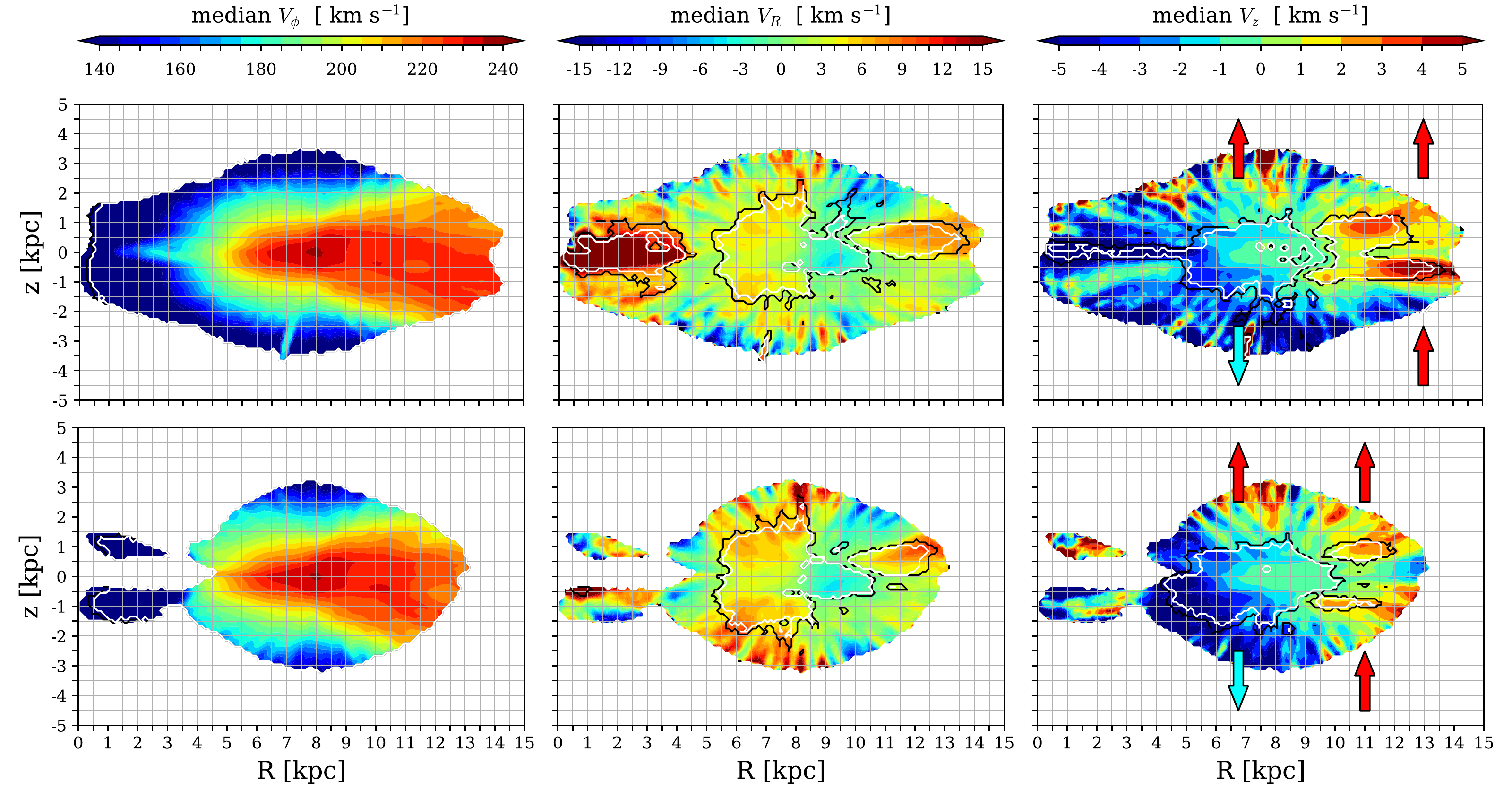}
	\caption{Maps of median values of each component of Galactocentric velocity as a function of ($R,z$), obtained using the distances derived with the {\tt StarHorse} code (top) and the ones by \citet{Paul2018} (bottom). The maps are shown in (0.1\,kpc)$^2$ pixels with a minimum of 50 stars. Each map has been computed by taking the mean of 100 Monte Carlo iterations. The contours show velocities with a significance larger than $2\sigma$ (black) and $3\sigma$ (white). The arrows in the right column indicate the direction of vertical velocity. These show signatures of a bending mode perturbation outside $R_0$ and a breathing mode inside.} \label{fig:newvel}
\end{figure*}
\section{Models}
\label{sec:Models}

In Section~\ref{sec:Radgrad} we compare our sample from {\it Gaia} DR2 to the results obtained with two different Galactic disc models with properties close to those of the Milky Way. The first one is selected from the suite of zoom-in simulations in the cosmological context presented by \citet{Martig2009,Martig2012} (Halo106). The radius was rescaled to 65\% of its original value so as to match the Milky Way's disc scale-length and rotation curve as described by \citet{Minchev13}, who used the same model for their chemo-dynamical model. We refer to this simulation hereafter as Model 1. 

The second model is an $N$-body simulation presented by \citet{Laporte2017, Laporte2018}, which considered the interaction of a Sagittarius-like dSph with the Milky Way. This simulation has quantitatively demonstrated that Sagittarius can simultaneously account for the distribution of outer disc structures seen in the anticenter \citep[][]{Newberg2002,PriceW2015,Bergemann2018,Sheffield2018} and most recently, the phase space spiral in ($z, V_z$) and ridges in the $V_\phi-R$ plane \citep{Laporte2018}. We use snapshots 690 and 648 of their L2 model and refer to it as Model 2. For Model 2 we rescale the distance to $80\%$. 
\section{6D phase space}
\label{sec:3dVel}

\subsection{Velocity maps in the $R-z$ plane}
\begin{figure*}
	\includegraphics[width=\textwidth]{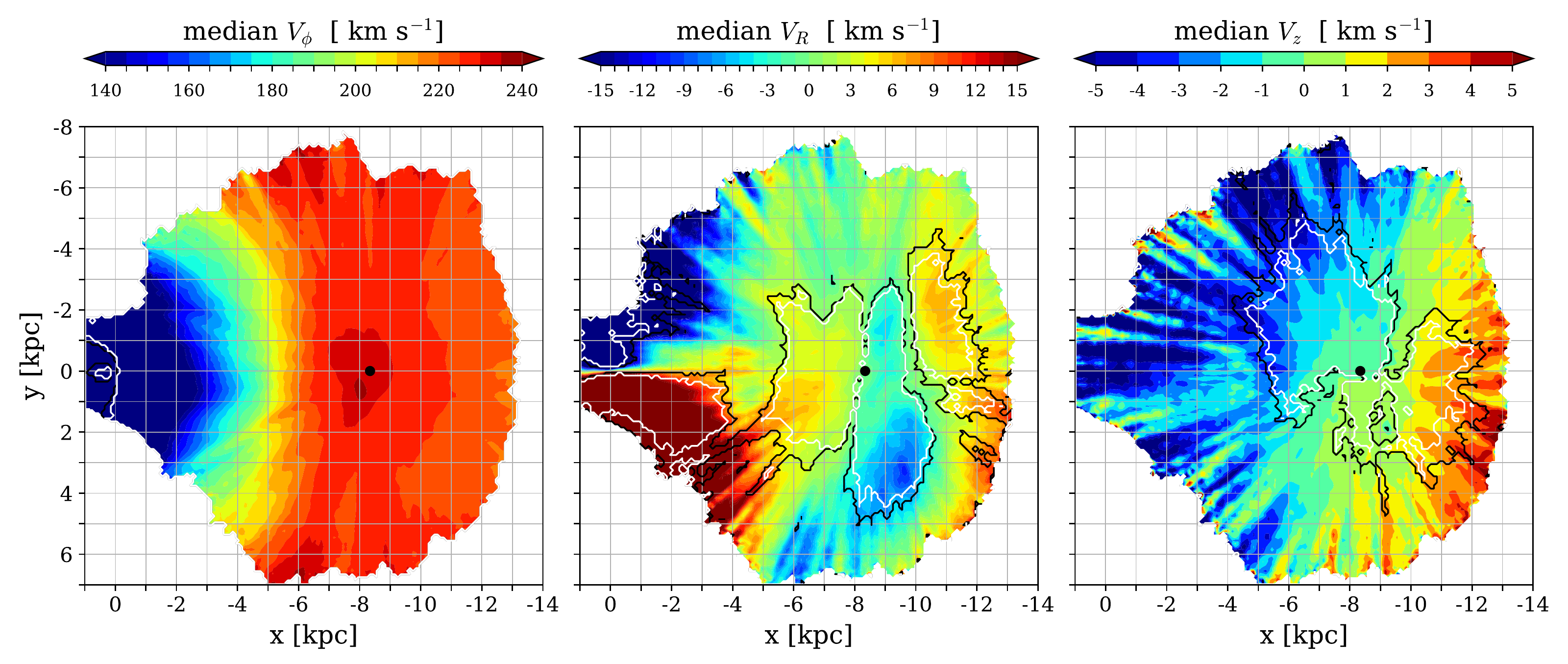}
	\caption{Same as Fig.\,\ref{fig:newvel} but for $x$-$y$ maps (face-on) using the {\tt StarHorse} distances. The Sun is located at $x,y=(-8.34,0)$\,kpc (black dot). The Galaxy rotates clockwise with its centre at $x,y=(0,0)$.} \label{fig:xyvel}
\end{figure*}
Figure \ref{fig:newvel} shows the median Galactocentric azimuthal ($V_{\phi}$), radial ($V_R$), and vertical ($V_z$) velocity fields obtained using the distances from {\tt StarHorse} (top) and the ones of \citet{Paul2018} (bottom) in the $R$-$z$ plane. The {\tt StarHorse} volume expands on the volume covered by the McMillan estimate because it has lower distance uncertainties at larger radii and it takes into account extinction, allowing for better estimates at lower latitude. The maps are shown in bins of (0.1\,kpc)$^2$, with a minimum of 50 stars per bin to avoid issues with low number statistics. Each map is computed by taking the mean of 100 realizations via Monte Carlo iterations. We do this by taking into account individual errors in distance, proper motion, line-of-sight velocity and drawing velocity values from a normal distribution centred on the originally estimated value. The black and white contours indicate significance larger than $2\sigma$ and $3\sigma$, respectively.

The $V_{\phi}$ maps (left column of Fig.\,\ref{fig:newvel}) show a clearly defined wedge shape for contours representing $V_\phi\lesssim220$~\,km\,s$^{-1}$ - as vertical distance increases at a given radius, we encounter slower rotating stars. This is related to the asymmetric drift effect, where stars with lower angular momentum reach outer radii at higher distance from the mid-plane.

The $V_R$ velocity fields (middle column of Fig.\,\ref{fig:newvel}) present a more complex variation with $R$ and $z$. For example, as evident from the 2 and 3$\sigma$ contours (overlaid black and white lines) in the top panels, we can see a negative radial gradient in $V_R$ at $R \lesssim 10 \,$kpc and $|z|\lesssim 2$\,kpc, and a positive one at larger radii (similar results are seen in the bottom panels, where the negative $V_z$ inside $R\sim5$\,kpc is probably due to the larger uncertainties above the 2 and 3$\sigma$ levels).
This is consistent with the results of \citetalias{Carrillo2018MNRAS} using {\it Gaia} DR1 and \citet{Katz2018} using {\it Gaia} DR2. 
Radial variation of $V_R$ appears similar but shifted to larger radius at larger $|z|$, possibly due to the asymmetric drift effect shifting higher eccentric orbits to larger $R$, similar to the shift in resonances described by \citet{Muhlbauer2003}.

\begin{figure*}
	\includegraphics[width=\textwidth]{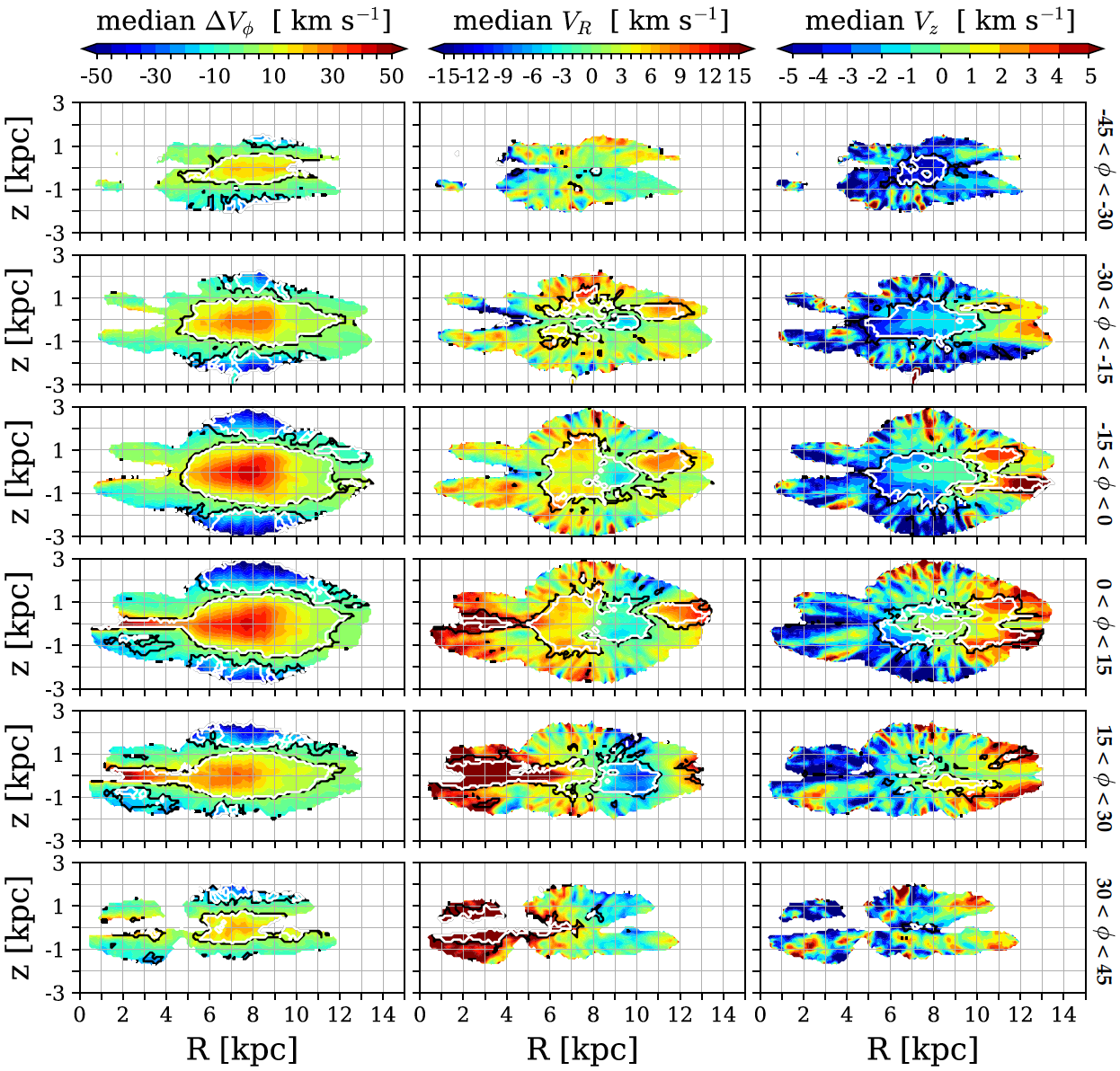}
	\caption{Same as Fig.\,\ref{fig:newvel} for different azimuth slices as indicated on the right. The first column shows the Galactocentric azimuthal velocity $V_{\phi}$ considering the effects of the asymmetric drift as described in the text. $V_{\phi}$ is shown to be the more symmetric component. In contrast, the radial velocity $V_R$ seems to strongly vary as a function of $\phi$. Changes in $V_z$ corresponding to different combinations of breathing and bending modes are shown in the right column.} \label{fig:slices}
\end{figure*}

The $V_z$ maps (right column of Fig.\,\ref{fig:newvel}) also show rich variations with radius and distance from the disc mid-plane. The arrows indicate bulk stellar motions at two different radii, showing a switch from a breathing at $R\approx6.5$\,kpc to a positive bending mode at $R\approx 13$\,kpc. A similar combination of modes was previously found in {\it Gaia} DR1 (\citetalias{Carrillo2018MNRAS}), where a negative bending was observed at $R\approx8.5$\,kpc.

Our data sample obtained with {\tt StarHorse} distances (top row) is not only more precise (Section~\ref{sec:Distances}) but also covers a larger volume than the sample using \cite{Paul2018} distances. We will, therefore, continue our analysis with the former dataset.

\subsection{Velocity maps in the $x-y$ plane}

In Fig.\,\ref{fig:xyvel} we present colour maps in the ($x\text{-}y$) plane for all velocity components, as indicated. The Sun is located at $x,y=(-8.34,0)$\,kpc, the Galactic centre is at $x,y=(0,0)$\,kpc, the Galaxy rotation is clockwise and there is no cut in $z$. This orientation is the same as in \cite{Katz2018}, Fig.~10.

The strong radial gradient in $V_\phi$ apparent in Fig.\,\ref{fig:newvel} is also seen in the left panel of Fig.\,\ref{fig:xyvel}. For $R\gtrsim 8$\,kpc some azimuthal variations may be present on the order of 15 \,km\,s$^{-1}$, while at $R\lesssim -4$\,kpc strong double peaks are seen. Away from the Sun, the drop in $V_\phi$ may be due to selection effects, where the density of stars close to the plane decreases with distance from the Sun.

The middle panel of Fig.\,\ref{fig:xyvel} presents the Galactocentric radial velocity, $V_R$. This shows similar structure to the top panels of Fig.~10 by \cite{Katz2018}, but the radial range covered is about twice as large. The larger volume covered in the inner disc is particularly useful to study the bar/bulge region. At $R \lesssim 4 \,$kpc with a significance of at least $3\sigma$ (white contours), we see a clear separation between stars lagging and stars ahead of the solar azimuth with negative and positive $V_R$, respectively. A more detailed analysis of this structure compared to our models will be done in Section~\ref{sec:Radgrad}.

The vertical velocity map in the right panel of Fig.\,\ref{fig:xyvel} shows an interesting variation close to the solar radius. Stars outside the solar radius exhibit upward stellar motion and downward motion inside. This pattern is consistent with the observed kinematic warp seen by \citet{Poggio2018}. The ridge of positive $V_z$ is, however, not exactly centred at $\phi=0$, as expected from this feature. \citet{Poggio2018} argued that this could be due to the Sun not being on the line of nodes. This is possible if the deviation is large, but probably we are just observing a combination of modes or it is just simply that the line of nodes has an inclination with respect to the plane as a function of radius. Such a non-straight line of nodes is consistent with recent results presented by \citet{Romero2018}, who used data from {\it Gaia} DR2 to study the global shape of the warp.
\subsection{Variations with Galactic azimuth}
\begin{figure*}
	\includegraphics[width=\textwidth]{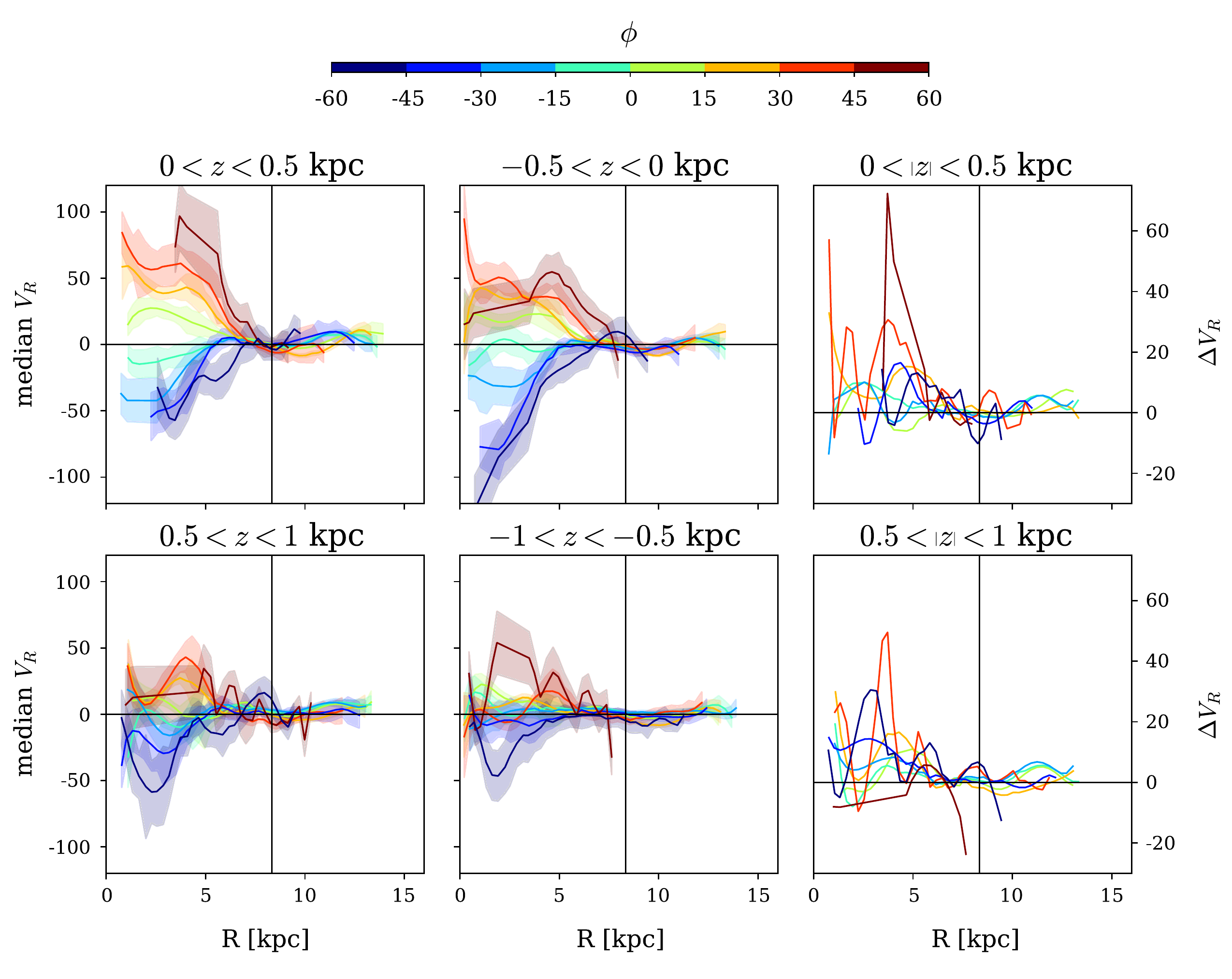}
	\caption{$V_R$ as a function of $R$ for different azimuth slices $\phi$ and heights from the plane $z$. The error bars correspond to the standard error of the median. The black solid lines represent $V_R=0$\,km\,s$^{-1}$ and the solar position. The left and middle column show the results obtained for the northern and southern hemisphere, respectively, while the right column shows the $V_R$ residuals between both Galactic hemispheres. The striking feature is the systematic shift seen in the inner disc, smoothly changing from positive $V_R$ at positive $\phi$ to negative $V_R$ at negative $\phi$.}
	\label{fig:GradientVr}
\end{figure*}

Figure\,\ref{fig:xyvel} showed that there exists significant azimuthal variations in all three velocity components, suggesting that some structure in the $R\text{-}z$ plane velocity maps (Fig.\,\ref{fig:newvel}) may have been washed out since we integrated over the entire azimuthal range.

To study variation in velocity structure with Galactic azimuth ($\phi$), we divide our sample in 15$^\circ$ slices in the range $-45^\circ<\phi<45^\circ$. The results are shown in Fig.\,\ref{fig:slices}. The left column shows the residual $\Delta V_{\phi}$ in (0.1\,kpc)$^2$ bins, computed by subtracting the mean of the median $V_{\phi}$ at all heights within the same $R$, from the median $V_{\phi}$ of each $R$-$z$ bin. This removes the Galactic rotation and the effect of the asymmetric drift seen in Fig.\,\ref{fig:newvel}, allowing to better see structure at a given radius. In the inner disc, inside $R\sim3$\,kpc, we see a peak in $\Delta V_{\phi}$ at $0^\circ<\phi<30^\circ$ and a second one just inside the solar radius at $R\sim 7$\,kpc. These peaks decrease at large and small azimuths, likely indicating the crossing of spiral arms.

The middle column of Fig.\,\ref{fig:slices} also shows strong azimuthal dependence in $V_R$ bulk motions. The strongest variations are found in the inner disc and close to the mid-plane, in agreement with the maps in Fig.\,\ref{fig:xyvel} (which is dominated by stars close to the plane). Inside $R_0$, we see symmetrical radial velocities across $z=0$. However, at $R>R_0$, $V_R$ is asymmetric across the mid-plane for $\phi<0^\circ$ but mostly symmetric at $\phi>0^\circ$. This is interesting because if the radial streaming motions were solely due to spiral structure, they should be symmetric across the Galactic plane. This suggests external perturbations, e.g., from the Sagittarius dwarf galaxy, are at play here. In the next Section we will study the azimuthal variations in more detail and compare them to our two models.

The $V_z$ maps displayed in the right column of Fig.\,\ref{fig:slices} agree with the combined $R\text{-}z$ map showing a combination of breathing and bending modes (Fig.\,\ref{fig:newvel}, right column). Nonetheless, some interesting azimuthal variations are seen at $-30^\circ<\phi<-15^\circ$ within $10<R<12$\,kpc where negative $V_z$ is seen at $z\sim-1$ and positive $V_z$ above, corresponding to a breathing mode. This contrasts with the strong positive $V_z$ seen at $0^\circ<\phi<15^\circ$ above and below the plane and the negative bending mode seen outside the solar radius in the range $-45^\circ<\phi<-30^\circ$. Both the breathing and the negative bending mode, however, suffer from large uncertainties (see white and black contours) probably due to a low number of stars at these azimuths.
\subsection{Radial velocity gradient}
\label{sec:Radgrad}

\begin{figure*}
	\includegraphics[scale=1.35]{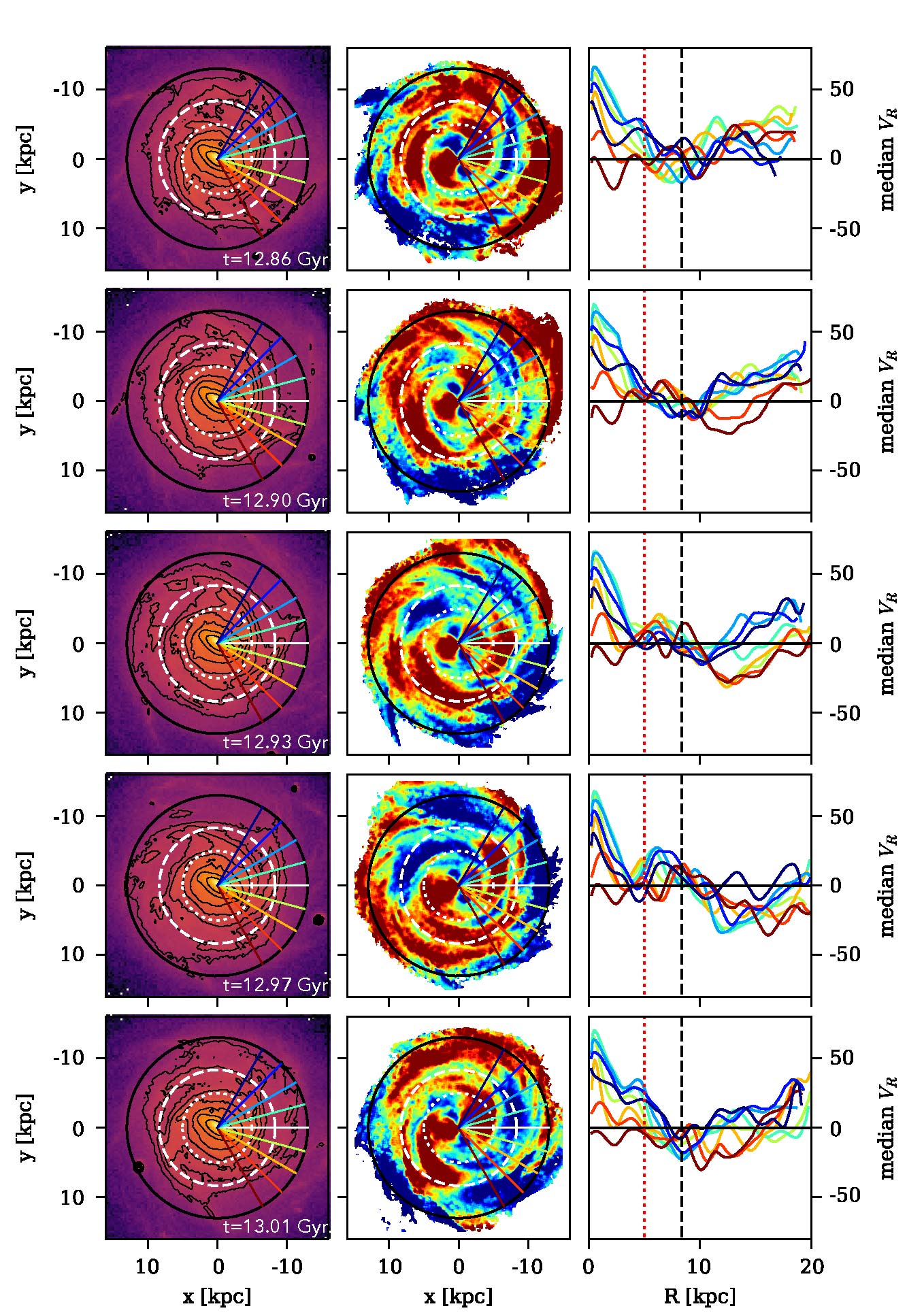}
	\caption{Snapshots of different evolutionary times obtained with Model 1, with the bar always oriented at $\phi=30 \,^{\circ}$ with respect to the solar azimuth. The left column shows density plots of the galaxy, the dashed white circle indicates the solar position and the dotted white circle (red line in right column) the radius of the bar region. The colour lines represent the same azimuthal ranges as in Fig.\,\ref{fig:GradientVr}. Face-on maps of the radial velocity are presented in the middle column. For comparison, the right column illustrates the same as in Fig.\,\ref{fig:GradientVr} within $-0.5<z<0.5$\,kpc but using the simulation.} \label{fig:densityplots}
\end{figure*}

\begin{figure*}
	\includegraphics[width=\textwidth]{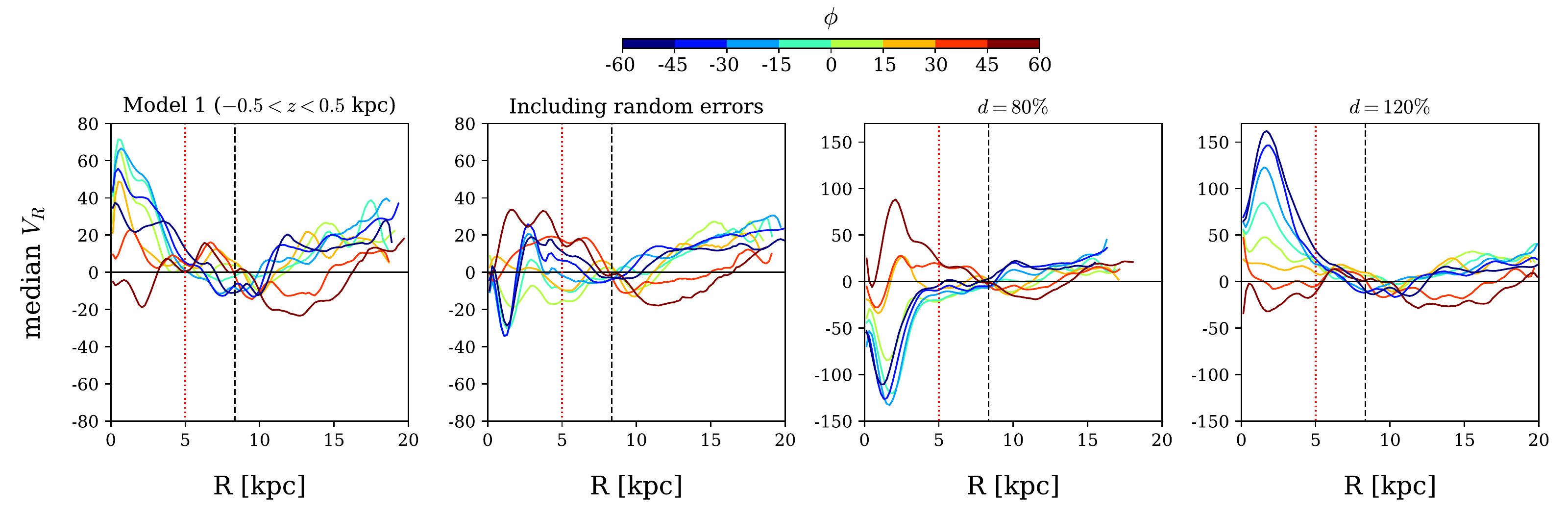}
	\caption{Model 1 $V_R$ as a function of Galactic radius for different azimuths (as colour-coded) at t=12.90 Gyr (1\textsuperscript{st} panel) and the effects of including random errors in $\mu$, $d$ and $V_{\mathrm{\,los}}$ (2\textsuperscript{nd} panel), as described in the text. The 3\textsuperscript{rd} and 4\textsuperscript{th} panels show the effect of recomputing $V_R$, while scaling the distance to $80\%$ and $120\%$, respectively. The patterns in the bar region (inside red dotted line) show drastic changes due to the included uncertainties. A reversal is seen for most azimuths when distances are affected by random motions (2\textsuperscript{nd} panel) and especially when they are systematically decreased (3\textsuperscript{rd} panel).} \label{fig:addederrors}
\end{figure*}

Here we investigate in more detail the variation of the $V_R$ radial gradient with Galactic azimuth and distance from the disc mid-plane. The $V_R$ map shown in Fig.\,\ref{fig:xyvel} is not very informative as the colorbar range of $\pm15$\,km\,s$^{-1}$ was chosen in order to show structure in the outer disc, which resulted in saturation in the inner one. Hence, in Fig.\,\ref{fig:GradientVr} we show the median $V_R$ as a function of $R$ for $15^\circ$ azimuthal slices in the range $-60^\circ<\phi<60^\circ$, as indicated by the colorbar and divide our sample between $0<\,\left |\, z\, \right |<\,0.5 $\,kpc (top) and $0.5<\left | z \right |< 1 $\,kpc (bottom). The left column shows the results obtained for stars above and the middle column for stars below the plane. The error bars correspond to the standard error of the median. The black solid lines represent the solar position and $V_R=0$\,km\,s$^{-1}$.

At distances closer to the disc mid-plane (top panels of Fig.\,\ref{fig:GradientVr}), variations with azimuth are on the order of 10\,km\,s$^{-1}$ near the solar radius, increasing outside $R_0$ to 30\,km\,s$^{-1}$ and especially toward the bulge region, where values of up to $\pm 100$\,km\,s$^{-1}$ are seen. In the inner disc, we observe a strong radial $V_R$ gradient, transitioning smoothly from $\approx +16$\,km\,s$^{-1}$\,kpc$^{-1}$ in a range between $30^\circ<\phi<45^\circ$ ahead of the Sun-Galactic centre line to $-16$\,km\,s$^{-1}$\,kpc$^{-1}$ in $-45^\circ<\phi<-30^\circ$ lagging the solar azimuth. In contrast, at larger distances from the disc plane (bottom panels), the radial velocity amplitude decreases to approximately half, while the trends remain similar.

The right column of Fig.\,\ref{fig:GradientVr} displays the $V_R$ residuals between the northern and southern Galactic hemispheres $ \Delta V_R=\left | V_{R,\,North} \right |-\left | V_{R,\,South} \right |$. In the top right panel the residuals show at positive azimuths larger $V_R$ above the Galactic plane and at negative azimuths larger $V_R$ below. This type of shearing is not observed for stars farther away from the mid-plane, where larger $V_R$ are seen at $z>0$ for most azimuthal bins.

We are interested in understanding what may cause the azimuthal gradient in $V_R$. Therefore, we will focus our attention mostly to the inner disc. It is remarkable that the fanning of $V_R$ towards the Galactic centre shifts systematically from $\sim-50$\,km\,s$^{-1}$ for $-30^\circ<\phi<-15^\circ$ (light blue curves in Fig.\,\ref{fig:GradientVr}) to $\sim +50$\,km\,s$^{-1}$ for $15^\circ<\phi<30^\circ$ (yellow curves in Fig.\,\ref{fig:GradientVr}). It should be noted that wavy behaviour is present for each azimuth, with $V_R$ peaks shifting systematically with azimuth, as expected for a spiral pattern.

\begin{figure*}
	\includegraphics[width=\textwidth]{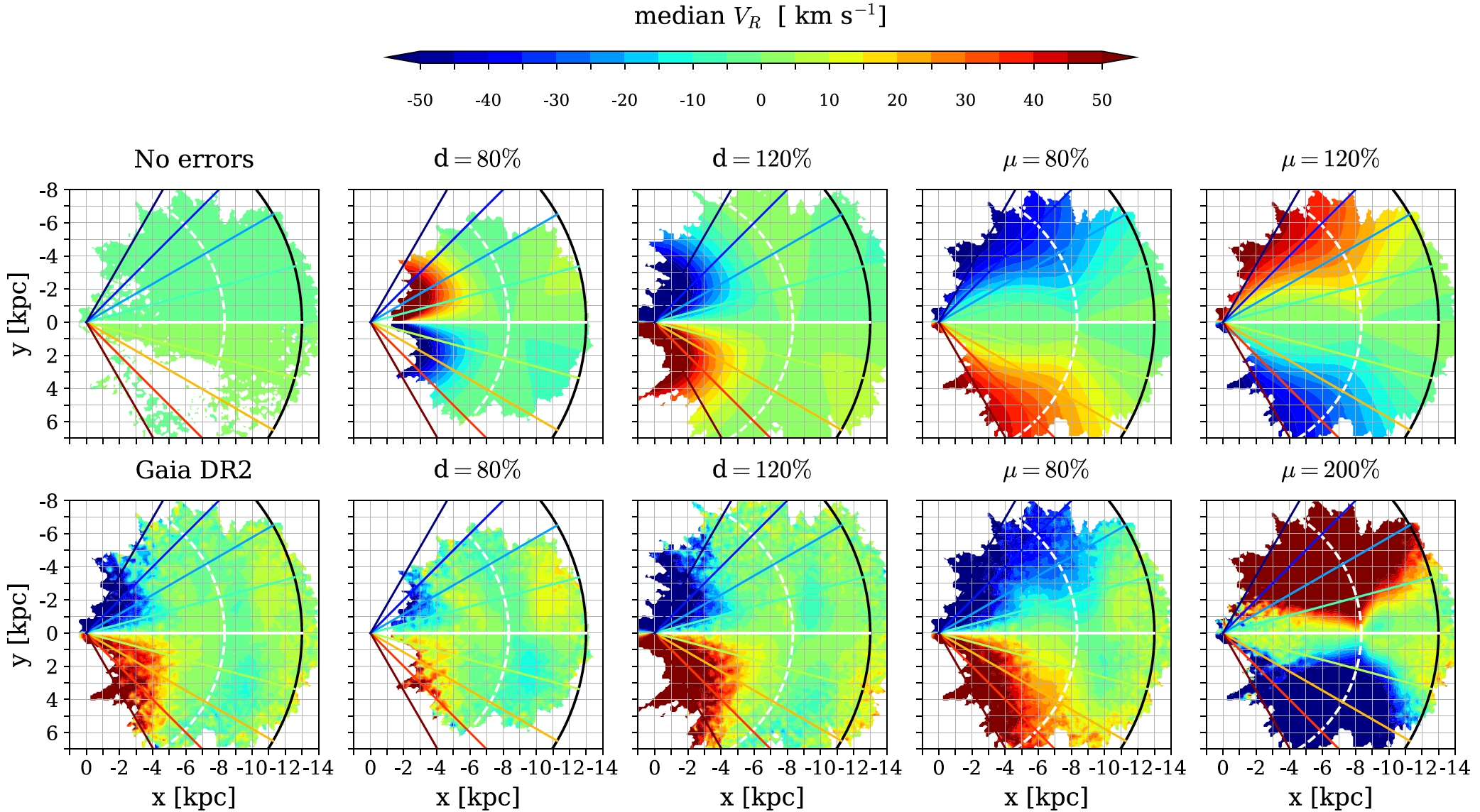}
	\caption{Top: Maps of median $V_R$ as a function of ($x,y$) for an idealized case of a Galaxy composed only of circular orbits obtained by setting $V_R=0$ and $V_{\phi}=240$\,km\,s$^{-1}$ in the data. Systematic errors are introduced as indicated on top of each panel and $V_R$ is recomputed. Bottom: Maps showing the effect of applying systematic errors to the data (the bottom left panel is the same as the middle one in Fig.\,\ref{fig:xyvel} but with a different $V_R$ range and excluding the $\sigma_{d}/d<0.2$ cut). As can be seen, high proper motion systematics invert the $V_R$ pattern but fail to reproduce the trends expected from Model 1 in the inner disc.} \label{fig:allsyst}
\end{figure*}

\subsubsection{The effect of a steady-state bar}

\begin{figure*}
	\includegraphics[width=\textwidth]{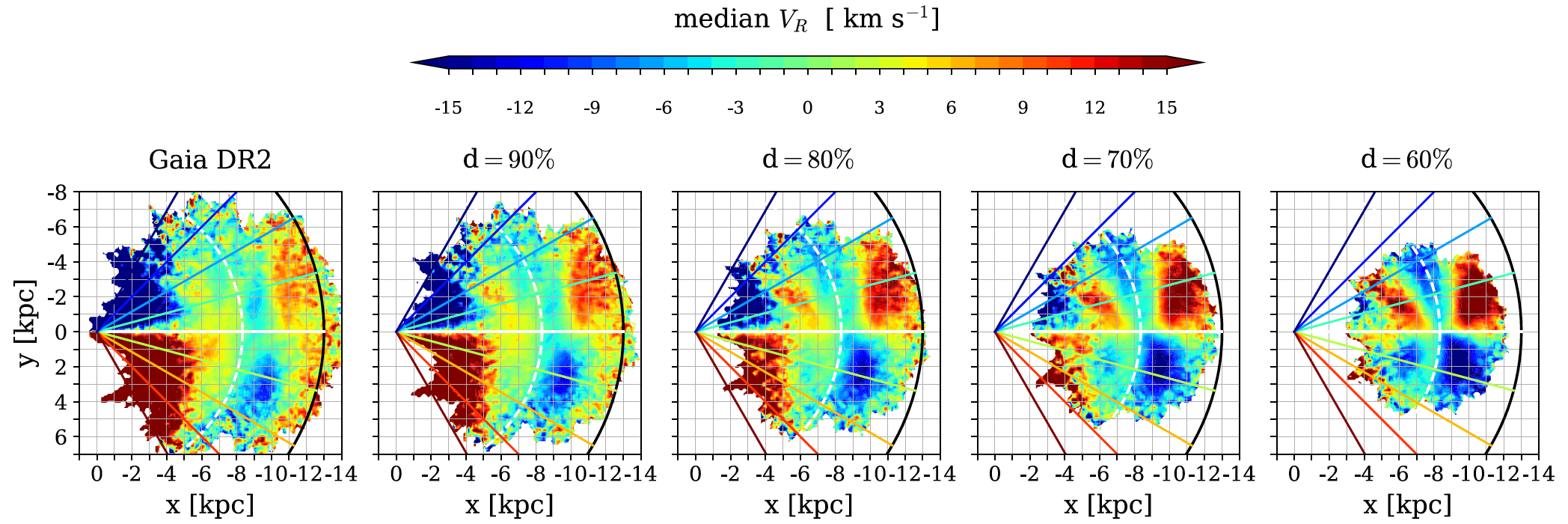}
	\caption{Exploring the effect of distance systematics applied to the data. Distances are multiplied by a factor of 0.9 through 0.6 (second through fifth panels, as indicated). As the systematics increase, the $V_R$ dipole in the inner disc decreases in amplitude, while outside the solar radius a dipole is created. Even at $40\%$ systematic error in distance, the inner disc trends observed in the first panel are not reversed as would be expected from the idealized case (second top panel of Fig.\,\ref{fig:allsyst}).} \label{fig:distsyst}
\end{figure*}

\begin{figure*}
	\includegraphics[width=\textwidth]{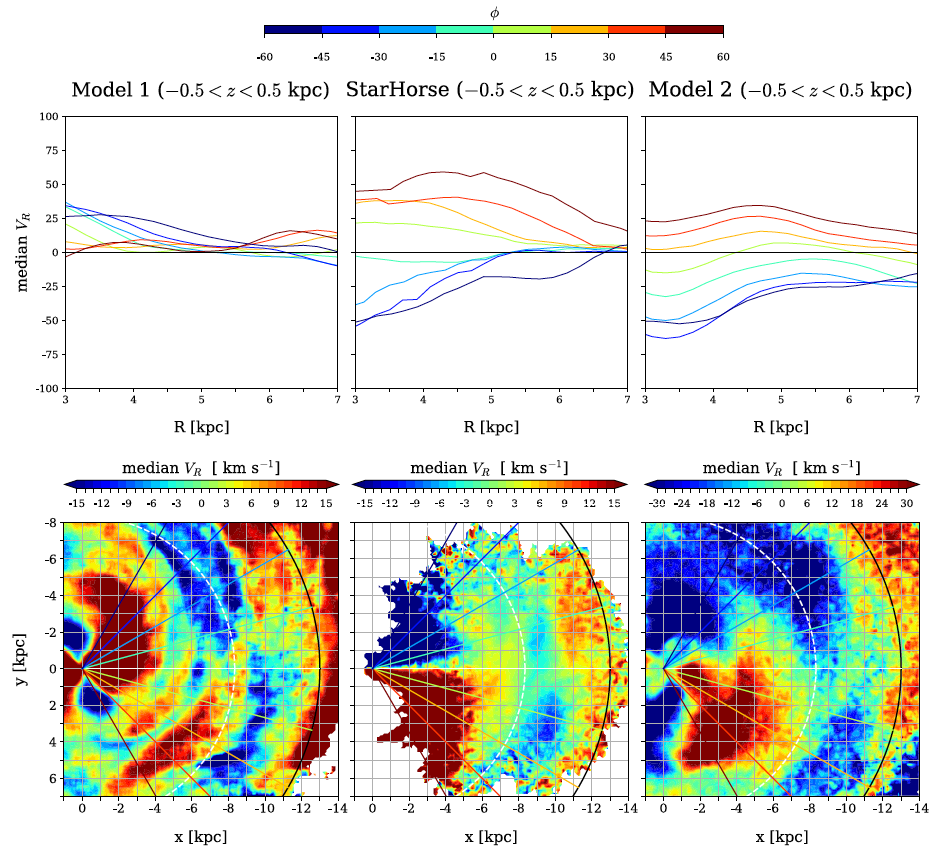}
	\caption{Comparison between data and models. Top: $V_R$ variation with $R$ for different azimuths resulting from Model 1 (left), data (middle), and Model 2 (right). Bottom: Corresponding face-on $V_R$ maps. Note the different colorbar range in the right bottom panel, compared to the left and middle ones. In both models, the Galactic bar was fixed at $\phi=30 \,^{\circ}$, roughly the angle that the Sun lags the bar. The trends in data are reproduced in the Model 2, which considers the interaction between the Milky Way and the Sagittarius dwarf galaxy.} \label{fig:chervinVr}
\end{figure*}

\begin{figure*}
	\includegraphics[width=\textwidth]{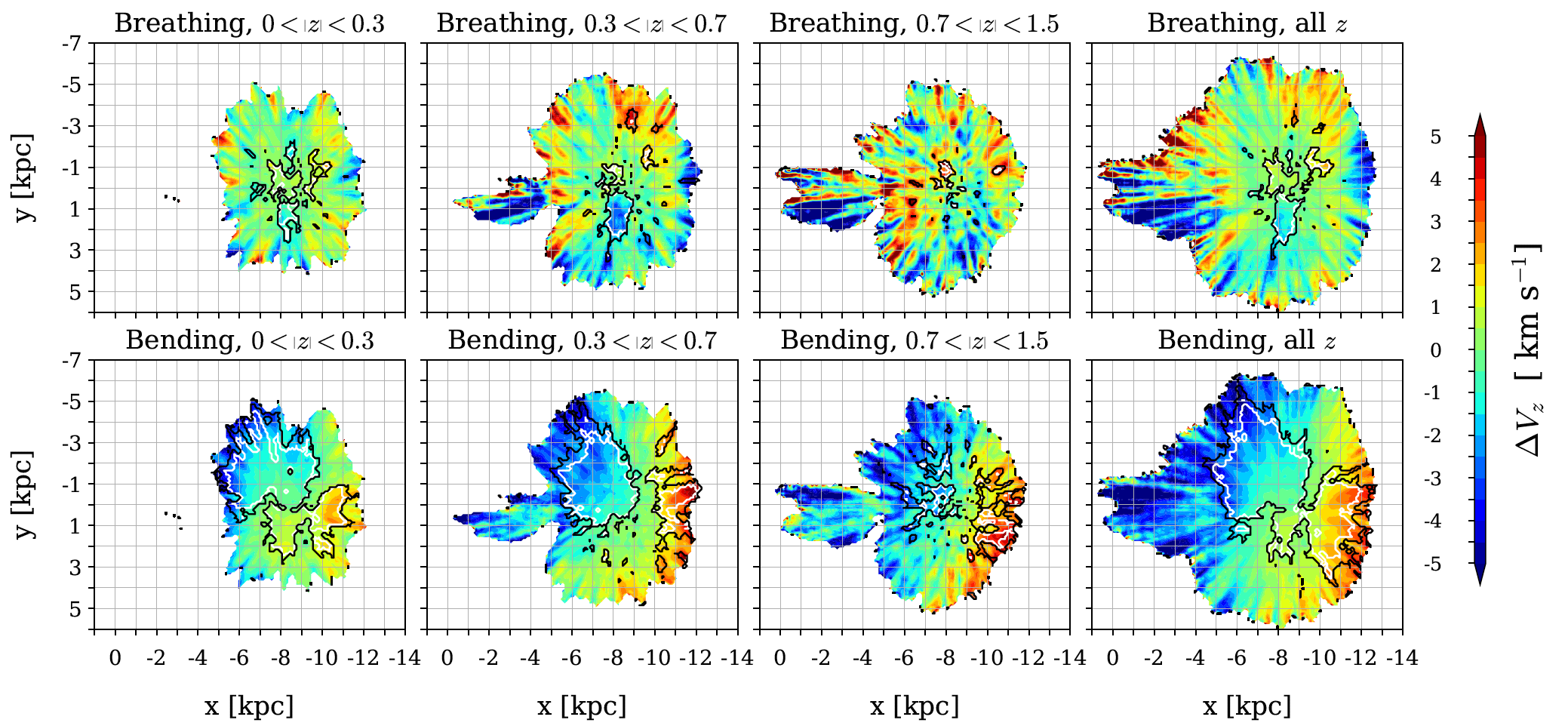}
	\caption{Breathing (top) and bending (bottom) modes estimated from the data for four different disc slices in vertical distance, $z$, as indicated on top of each panel (rightmost panels shows all $z$).} \label{fig:Breathbend}
\end{figure*}

\begin{figure*}
	\includegraphics[scale=.47]{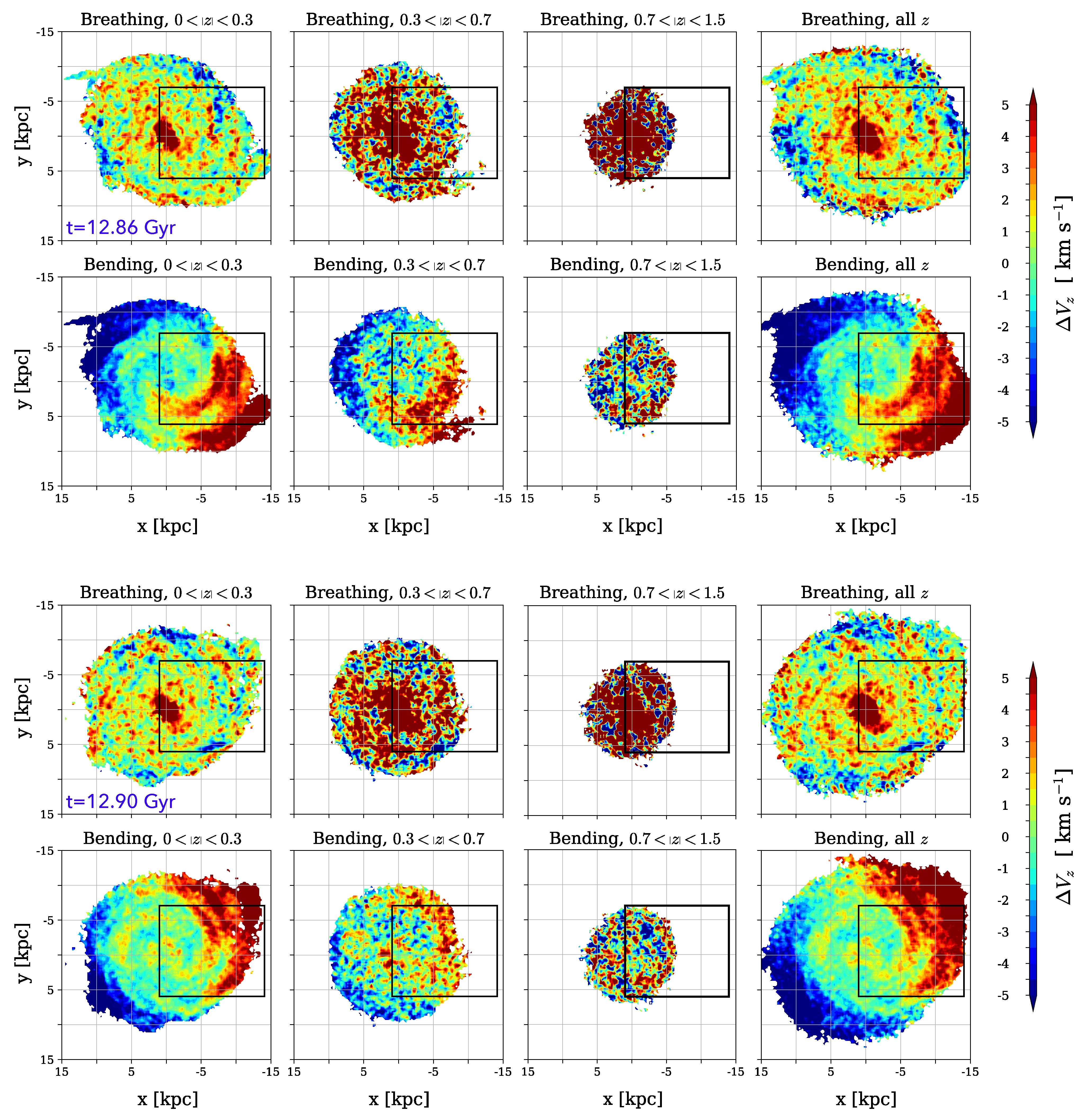}
	\caption{Same as Fig.\,\ref{fig:Breathbend} for the Model 1 at two different evolutionary stages. The black squares indicate the data volume. While the top panel exhibits a similar behaviour as the data, the bottom panel shows at a short elapsed time the opposite behaviour.} \label{fig:Breathbendmodel}
\end{figure*}

\begin{figure*}
	\includegraphics[width=.9\textwidth]{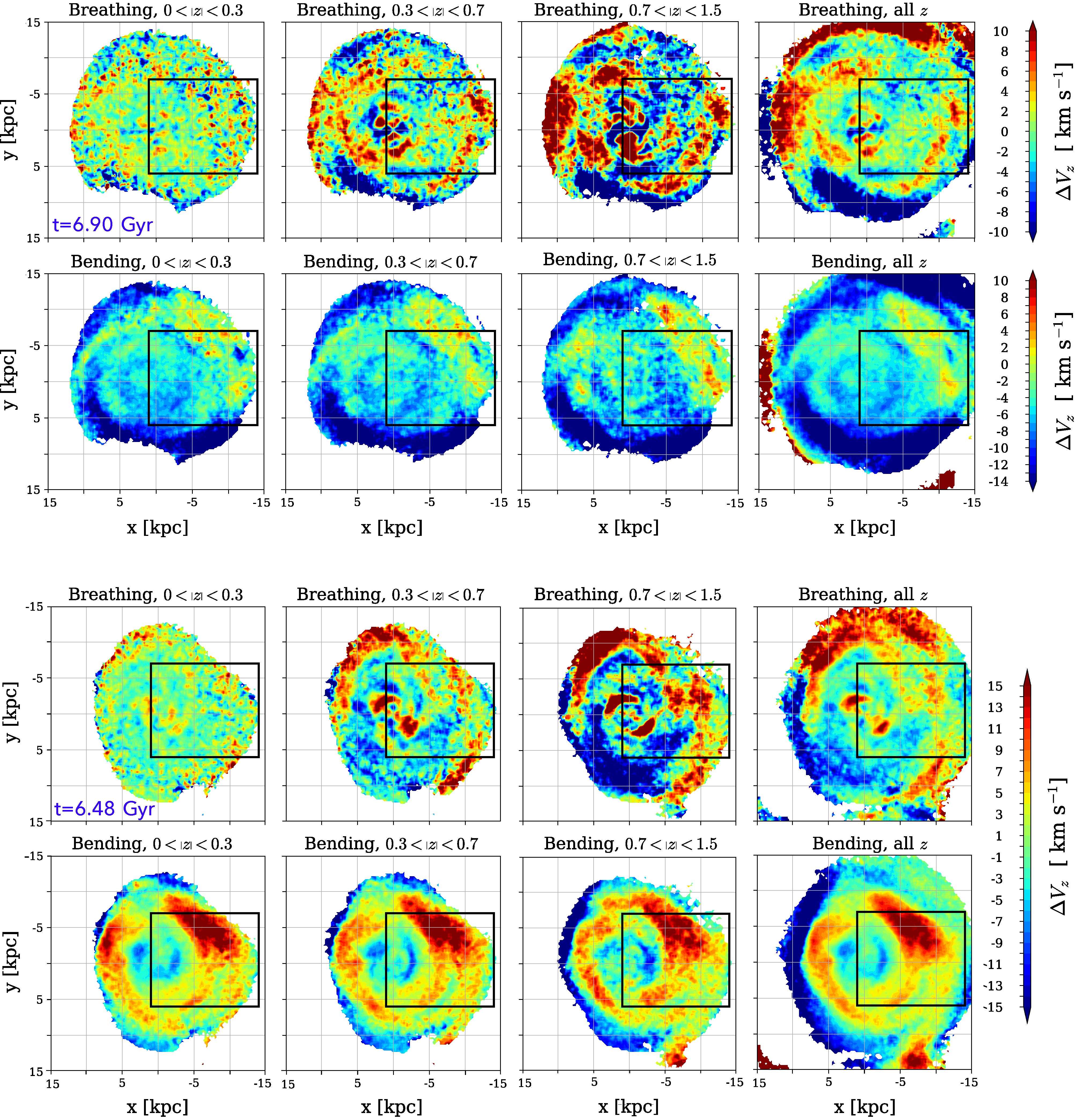}
	\caption{Same as Fig.\,\ref{fig:Breathbendmodel} for the Model 2. Here the mismatch with the data occurs at an earlier evolutionary stage (bottom panels), in contrast to Model 1. The constant vertical variations with time, make the understanding of the $V_z$ patterns quite complex.} \label{fig:Breathbend_chervin}
\end{figure*}

In order to understand the origins of such a pattern we now use the models described in Section~\ref{sec:Models}. The left column of Fig.\,\ref{fig:densityplots} shows face-on density maps for 5 different snapshots of Model 1, separated by $37.5$ Myr. During this time interval there are no significant mergers disturbing the disc. The dashed white circle represents the solar neighbourhood, the dotted white circle the bar region, and the colour lines indicate the $\phi$ angles as in Fig.\,\ref{fig:GradientVr}. For this model, we fixed the Galactic bar at roughly the angle that the Sun lags the bar ($\phi=30 \,^{\circ}$), while we let the galaxy rotate clockwise. The constant density contour levels are spaced by $10\%$ from 90 to $30\%$. The $V_R$ maps due to the interaction of the spiral arms with the bar are shown in the middle column of Fig.\,\ref{fig:densityplots}. The changes between different evolutionary stages are more evident outside the solar radius. This can be seen better in the right column, showing $V_R$ as a function of radius for different azimuths (as in Fig.\,\ref{fig:GradientVr}). In the bar region (inside the red dotted line), however, it can be seen that, independently of time, $V_R$ exhibits a similar pattern with velocity decreasing with increasing $\phi$. This pattern remains stable for later evolutionary stages and is the expected velocity structure due to the bar rotation. These negative/positive streaming at the leading/trailing bar side can also be seen in the Auriga simulations (e.g. \citealp{Grand2016}, Figure 1), as well as test particle models (\citealp{Monari2016}, Figure 4). It is therefore very interesting, that the data shown in Fig.\,\ref{fig:GradientVr} shows exactly opposite trends inside $R\sim5$\,kpc.

\subsubsection{The effect of systematics}
\label{sec:syst}

To see if this discrepancy can be explained by uncertainties in the data, we add random errors to one of the snapshots (t=12.90 Gyr). We use the data typical errors in proper motion $\sigma_{\mu}= 0.07$ mas\, yr$^{-1}$ and line-of-sight velocity $\sigma_{\mathrm{\,los}}=1.8$\,km\,s$^{-1}$ together with the maximum accepted distance error of $\sigma_d=20\%$. The results can be seen in the second panel of Fig.\,\ref{fig:addederrors}. Compared to the snapshot without errors in the first panel, the inclusion of random errors is enough to affect the pattern inside the solar radius, but the systematic shift as a function of azimuth is not as clear as in the data. The third and fourth panels of Fig.\,\ref{fig:addederrors} show the effect of possible systematics in our distances. For this, we multiply the distance by 0.8 and 1.2 and recompute $V_R$ (this is a simplified attempt to account for the problem as in reality the parallax shift might vary a lot with position and even colour of stars creating larger uncertainties). In the fourth panel, scaling the distance to $120\%$ increases the amplitude of the original pattern. In contrast, scaling the distance to $80\%$ (third panel) modifies the velocity amplitudes and inverts the trends for different azimuth systematically similar to the data.

To further study the effect of systematics, we create an idealized experiment of a Galaxy composed only of circular orbits by setting $V_R=\,0$ and $V_{\phi}=240$\,km\,s$^{-1}$ for all the stars. We continue by introducing systematic errors in distance and proper motion and recomputing the velocities. The velocity maps shown in Fig.\,\ref{fig:allsyst} and Fig.\,\ref{fig:distsyst} were computed with the same cuts as described in Section~\ref{sec:selection}, but do not include the $\sigma_{d}/d<0.2$ cut, as this would vary for each map. The top panels of Fig.\,\ref{fig:allsyst} show the effects of including systematic errors of $20\%$. As can be seen, distance and proper motion systematics, both create a systematic shift as a function of $\phi$. 
While proper motion systematics are unlikely the main feature creating the $V_R$ dipole due to the $200\%$ systematic error needed to invert the pattern and the affected area not being in the bar region, a $20\%$ systematic error in distance creates a trend that is quite similar to the dipole observed in the data at similar amplitudes. This would imply that our distances could be overestimated. Such systematic errors could be the effect of the assumed parallax zero-point shift. However, {\tt StarHorse} assumes a parallax shift of 0.05 mas, in agreement with recent works (e.g. \citealp{Zinn2018}; \citealp{Graczyk2019}; \citealp{Schonrich2019}). If the distances were overestimated this would require a larger parallax shift. Furthermore, applying these errors to the data (bottom panels of Fig.\,\ref{fig:allsyst}) shows that the pattern still remains even with a $20\%$ systematic error in distance. If systematics are indeed the main factor creating this $V_R$ trend, there should be a systematic distance or proper motion factor that inverts the pattern to match that expected from a typical bar model (Model 1). In Fig.\,\ref{fig:distsyst}, we further study the effect of distance systematics applied to the data. The maps show how a dipole is gradually created in the outer disc with decreasing distance and the one in the inner disc decreases in amplitude. The trends in the inner disc are never reversed (as seen in the second top panel of Fig.\,\ref{fig:allsyst}), even if we consider a $40\%$ systematic error. This indicates that the observed systematic shift in $V_R$ as a function of $\phi$ could be a real kinematic feature.
\subsubsection{The effect of Sagittarius}

We next study Model 2 to find out how an external perturbation from the Sagittarius dwarf galaxy modifies the inner disc velocity field.

The top row of Fig.\,\ref{fig:chervinVr} shows the variation of radial velocity with Galactic radius for different azimuthal slices (similar to Fig.\,\ref{fig:GradientVr} but for the range $3<R<7$\,kpc) for Model 1 (left), {\it Gaia} DR2 (middle), and Model 2 (right). The bottom row shows the same samples but as maps in the $x\text{-}y$ plane. 

We now see more clearly that Model 1, which lacks a significant external perturbation at the time period we consider, cannot reproduce the Galactocentric radial velocity structure in the bar vicinity. While the data reveal a region of high negative bulk velocity at azimuths lagging the bar and high positive velocity ahead of the bar, Model 1 showed the opposite trends, which we argued earlier are typical for a steady-state bar. This indicates that the Galactic bar and spiral arms cannot reproduce the azimuthal $V_R$ gradient alone.

The right column in Fig.\,\ref{fig:chervinVr}, however, shows that the disc perturbed by Sagittarius matches the data in the trends, including the concave behaviour of the largest azimuthal bins (orange and red curves). In other words, the radial velocity field streaming motions on each side of the closer bar half are completely reversed compared to a steady-state model. We note that these trends in the inner disc can also be caused by distance systematics as discussed in Section~\ref{sec:syst} but assuming that is not the case, we find a good match in Model 2. 

Looking at the bottom panels of Fig.\,\ref{fig:chervinVr}, we find a remarkable match of the entire $V_R$ field between {\it Gaia} DR2 and Model 2, all the way out to 14\,kpc. This includes the bridge between 8 and 10\,kpc with $V_R\approx 0$\,km\,s$^{-1}$, also recently showed for the same simulation by \citet{Laporte2018}, but note that here we have rescaled radius down to $80\%$. At a radius of about 14\,kpc we find an arch of outward streaming motions in data and both models, although a better match again results from Model 2, except for the amplitude being larger by a factor of $\sim2$.
\subsubsection{A recently formed/evolved bar}
The morphology of the Model 2 bar region at this time output can be seen in the bottom left panel of Figure 4 by \cite{Laporte2018}. Note that the bar formation in this model started about 1 Gyr ago, triggered by the impact. This time is not sufficient for the disc to respond fully to the bar perturbation. \citet{Minchev2010} showed that, up to 2\,Gyr after bar formation the $x_1(1)$ and $x_1(2)$ orbits at the bar OLR precessed in the bar reference frame thus giving rise to two low-velocity streams consistent with Pleiades and Coma Berenices or Pleiades and Sirius, depending on the time since bar formation (in addition to Hercules). This is unlike the expected behaviour of a steady-state bar, where these two orbital families are fixed in the bar reference frame along the bar major axis ($x_1(1)$) or perpendicular to it ($x_1(2)$). It is possible that such effects are present in the bar region as well, which could explain the inversion in the radial velocity field. A closer inspection of the central regions in both models shows that the inner $1\text{-}2$\,kpc are consistent in the expected trends from a steady-state bar. However, at radii nearing the bar end ($\sim4$\,kpc in Model 1 and $\sim3.5$\,kpc in Model 2) Model 1 never reproduces the large area of positive motions needed to match the data, for any time output.

It is also interesting to note that the bar farther half does not show the same $V_R$ trends as the near one for Model 2, further strengthening our conclusion that the bar is time evolving. This can be contrasted to the near-perfect symmetry found in the central 2\,kpc of the Model 1 disc. Future work will be dedicated to more detailed analyses.

\subsection{Breathing and bending modes}\label{sec:Vz_pattern}
In this Section, we focus our attention to the vertical velocity distribution, which has a complex structure with origins still under debate. In particular, we study the $V_z$ distribution by computing the breathing and bending modes in our sample and comparing these to our Models. Disentangling the observed mode, will helps us to put some constraints on the origins of the perturbation, as bending modes are attributed mainly to external perturbations and breathing modes are mainly induced by internal bar or spiral perturbations.

We define a breathing and bending mode based on the convention of \citet{Widrow2014}:
\begin{equation*}
\Delta V_{z,\,Breathing}=\left ( V_{z,\,North}-V_{z,\,South} \right )
\end{equation*}
and
\begin{equation*}
\Delta V_{z,\,Bending}=\frac{1}{2}\left ( V_{z,\,North}+V_{z,\,South} \right )
\end{equation*}

We note that the structure resulting from the application of the above equations to the data and models will not necessarily signify breathing and bending modes, which need self gravity to propagate, but could be the result of phase mixing due to an external perturbation (such as Sagittarius) as discussed by \citet{delaVega2015}.

Figure \ref{fig:Breathbend} displays the breathing (top) and bending (bottom) face-on maps for different $z$ ranges, obtained from 100 Monte Carlo iterations as in Fig.\,\ref{fig:xyvel}. The breathing pattern observed close to the Galactic plane ($0<\left | z \right |< 0.3 $\,kpc, top 1\textsuperscript{st} column) is consistent with a median $\Delta V_z\approx 0$. This indicates bulk vertical motions pointing in the same direction and with similar amplitude above and below the plane and thus no breathing mode. This is in agreement with the results found by \citet{Bennett2019} within $\left | z \right |< 1$\,kpc with an amplitude $< 1 $\,km\,s$^{-1}$ (half the amplitude compared to the breathing mode definition used in this work). However, at higher $z$ (top 2\textsuperscript{nd} and 3\textsuperscript{rd} column), the breathing mode seems to take shape, with mostly positive velocities at larger distances. This becomes more evident on the top 4\textsuperscript{th} column showing amplitudes beyond $5$\,km\,s$^{-1}$ in the overall map.

Unlike the breathing mode, all the bending mode patterns in the bottom panels of Figure \ref{fig:Breathbend} display the same trend, showing clear signatures of negative velocities at $R \lesssim 9$\,kpc and positive outside. These patterns are quite similar to the $V_z$ map shown in the right panel of Fig.\,\ref{fig:xyvel}, implying that the bending mode is the dominant mode in our sample.

In Fig.\,\ref{fig:Breathbendmodel} we compare two snapshots from Model 1 with our data. The data volume is illustrated by the black squares on each panel. Aside from the breathing mode at $0.7<\left | z \right |< 1.5 $\,kpc, the breathing and bending modes at t=12.86 Gyr seem to follow a similar pattern as the data. However, just $37.5$\,Myr later at t=12.90\,Gyr, the bending pattern strongly changes. Similar results are seen in Fig.\,\ref{fig:Breathbend_chervin} with Model 2. At t= 6.9 Gyr, after virial radius crossing and at all heights, we observe patterns which follow similar trends as the data. In contrast, at t=6.48 Gyr, roughly 0.42 Gyr before, the disc is strongly bent and does not match the observations. These strong changes show that understanding the origin of the $V_z$ patterns can be rather complex. Beyond the scope of this work, a detailed analysis of individual perturbations of the Galactic bar, the spirals and external perturbations is required to understand the observed vertical motions.

\section{Summary and Conclusions} \label{sec:Concl} 
In this work we have used the high accuracy proper motions and line-of-sight velocity obtained from {\it Gaia} DR2 combined with two sets of distances to study the three-dimensional velocity distribution of stars across a large portion of the Milky Way disc. The distances were obtained via Bayesian inference with two different approaches. Using the observed {\it Gaia} parallax $\varpi$ and the radial velocity spectrometer magnitude $G_{RVS}$ as priors \citep{Paul2018} and using the {\tt StarHorse} code, which combines the {\it Gaia} astrometric information with multiband photometric information and a number of Galactic priors \citep{Anders2019}. In Section~\ref{sec:Distances}, we demonstrated how both estimates improve greatly the volume and accuracy of stellar distance compared to the na\"{i}ve use of the inverse parallax as distance estimator: the observed volume was roughly doubled compared to previous authors using the inverse parallax and the mean uncertainties at $R>3$\,kpc are at the level of $\sim20\%$ instead of $\sim60\%$ when using the inverse parallax (see left panel of Fig.\,\ref{fig:distcomp}).

We summarize our main results as follows:
\begin{itemize}
	\item In both subsamples we identified asymmetries in the stellar velocity distribution. We studied the $V_R$ velocity distribution by dividing our sample in different azimuthal slices and heights from the plane. In the outer bulge/inner disc, we found variations up to $\pm 100$\,km\,s$^{-1}$.

	\item Inside the solar circle, we identified a strong radial $V_R$ gradient, transitioning smoothly from $-16$\,km\,s$^{-1}$\,kpc$^{-1}$ at an azimuth of $-45^\circ<\phi<-30^\circ$ lagging the Sun-Galactic centre line to $+16$\,km\,s$^{-1}$\,kpc$^{-1}$ at an azimuth of $30^\circ<\phi<45^\circ$ ahead of the solar azimuth.

	\item To understand the origins of such a gradient, we analysed the effect of introducing systematic errors in distance and proper motion. We found that distance systematics can create a similar pattern in the idealized case of a Galaxy composed only of circular orbits. However, when applied to the data, even correction for a large systematic shift did not remove the observed gradient in the inner disc.

	\item Assuming the gradient to be a real kinematic feature, we compared the data with two Milky Way models: a zoom-in simulation in the cosmological context, lacking recent massive external perturbers (Model 1) and an $N$-body simulation, which considers a Sagittarius-like interaction with the Milky Way (model 2). We found, that we were not able to reproduce the observed $V_R$ azimuthal gradient by considering mainly the internal effects of the Galactic bar and spiral arms. Instead, we demonstrated that introducing a major perturbation, such as the impact of Sagittarius, could reproduce a similar velocity field as in the observations (see Fig.\,\ref{fig:chervinVr}).
	
	\item The velocity field can be reversed in the bar region only if the bar were currently forming or a previously existing one has been recently strongly perturbed, both of which could have been caused by Sagittarius. This suggest that the reversed $V_R$ field in the inner disc is a strongly time-dependent phenomenon, as a steady-state bar must exhibit the $V_R$ streaming of Model\,1. As no external perturbations on the scale of Sagittarius are expected in the next several\,Gyr (except for the Sagittarius itself, who is constantly losing mass), we predict that the $V_R$ field should reverse in the next $1\text{-}2$\,Gyr to match Model\,1. 

	\item Consistent with the results first shown by \citetalias{Carrillo2018MNRAS}, we confirmed that the vertical velocity structure is a combination of vertical modes with a breathing mode inside the solar circle ($R \approx 6.5$\,kpc) and a bending mode outside ($R \approx 13$\,kpc). However, as \citet{Katz2018} previously showed, the bending mode differed from the one found at $R \approx 8.5$\,kpc by \citetalias{Carrillo2018MNRAS}. With a larger dataset expanding to larger distances, we observed a bending mode with positive $V_z$ instead of negative. 

	\item In the $x\text{-}y$ plane, we showed that, although the amplitude varies, the structure of the breathing and bending modes does not show any significant variations with distance from the Galactic plane. We showed that in both our models the breathing and bending patterns can drastically change in short time intervals, showing the complexity of understanding the origin of vertical perturbations. Thus, further and more detailed modelling is necessary to disentangle how internal and external perturbers individually affect the vertical velocity distribution in the Milky Way.
\end{itemize}
\section*{Acknowledgements}
I. Carrillo is grateful to E. Poggio for discussions concerning the Galactic warp.

This work has made use of data from the European Space Agency (ESA) mission
{\it Gaia} (\url{https://www.cosmos.esa.int/gaia}), processed by the {\it Gaia}
Data Processing and Analysis Consortium (DPAC,
\url{https://www.cosmos.esa.int/web/gaia/dpac/consortium}). Funding for the DPAC
has been provided by national institutions, in particular the institutions
participating in the {\it Gaia} Multilateral Agreement.

%%%%%%%%%%%%%%%%%%%%%%%%%%%%%%%%%%%%%%%%%%%%%%%%%%

%%%%%%%%%%%%%%%%%%%% REFERENCES %%%%%%%%%%%%%%%%%%

% The best way to enter references is to use BibTeX:

\bibliographystyle{mnras}
\bibliography{mine2} % if your bibtex file is called example.bib

% Alternatively you could enter them by hand, like this:
% This method is tedious and prone to error if you have lots of references
%\begin{thebibliography}{99}
%\bibitem[\protect\citeauthoryear{Author}{2012}]{Author2012}
%Author A.~N., 2013, Journal of Improbable Astronomy, 1, 1
%\bibitem[\protect\citeauthoryear{Others}{2013}]{Others2013}
%Others S., 2012, Journal of Interesting Stuff, 17, 198
%\end{thebibliography}

%%%%%%%%%%%%%%%%%%%%%%%%%%%%%%%%%%%%%%%%%%%%%%%%%%

%%%%%%%%%%%%%%%%% APPENDICES %%%%%%%%%%%%%%%%%%%%%

%\appendix

%If you want to present additional material which would interrupt the flow of the main paper,
%it can be placed in an Appendix which appears after the list of references.

%%%%%%%%%%%%%%%%%%%%%%%%%%%%%%%%%%%%%%%%%%%%%%%%%%

% Don't change these lines
\bsp	% typesetting comment
\label{lastpage}
\end{document}